# Experimental study on the elderly pedestrians passing through bottlenecks

Xiangxia Ren [a], Jun Zhang [a*], Shuchao Cao [b], Weiguo Song [a]

[a] State Key Laboratory of Fire Science, University of Science and Technology of China, Hefei 230027, People's Republic of China

[b] School of Automotive and Traffic Engineering, Jiangsu University, Zhenjiang 212013, People's Republic of China

**Abstract**

The aging of population is a social phenomenon in the world. In this study, series of controlled experiments were performed to investigate the movement characteristics of elderly pedestrians passing through bottlenecks. Similar self-organized phenomena to that of the young like 'zipper effect' and 'lane formation' are observed. The highest local density that appears in front of the exit in the experiment is beyond 4 m$^{-2}$. The time lapse between two consecutive pedestrians of the elderly is about 0.2 s longer than that of the young when the width is 0.5 m and 0.7 m. Besides, the difference decreases with the increasing exit width. A linear relation between flow and bottleneck width is obtained, while the flow of elders is lower than that of the young under the same width of exits. The waiting time of pedestrians can be divided into two stages when the bottleneck width is narrower than 0.8 m. The findings in this study will benefit the evacuation guidance as well as the facility design for elders.

## 1. Introduction

Bottleneck is a common geometry in most of facilities which restricts pedestrian's movement [1,2]. The study on pedestrian flow through bottlenecks attracted the attention of researchers and several experiments have been carried out under laboratory conditions to study the influence of the geometry, light intensity, competitiveness and initial distribution of pedestrians on the bottleneck flow [3–11]. It is found that the distance between the lanes in the bottleneck is independent of the exit width [3] but the process of the lane formation and the congestion degree are affected by the bottleneck width [4,12,13]. The peak density locates in front of the bottleneck [12,14] and the capacity increases with the additional lanes developing which leads to a stepwise increment [15]. The flow increases with the increasing bottleneck width firstly and it oscillates around a value when the bottleneck width is large [4,12]. The time lapses between two consecutive pedestrians are symmetrical with distinct maximum at the centre and become flatter with the increase of the bottleneck width [16]. Besides, the cumulative distribution of the time headway which can be used to describe the intermittent flow and the probability of finding a long-lasting clog in case of blockages was analysed for three degrees of competition but only two widths in [17]. The time to the target versus the distance to the target was used to exhibit the waiting time [18,19]. The steep degree of the tracked position line indicates the speed and congestion of the progress.

* Corresponding author: junz@ustc.edu.cn

However, most of these results are obtained from experiments with university students or young attendees, without considering elderly pedestrians. It is known that the ageing of the population is a social phenomenon in the world. The elderly pedestrians as a special group show different movement characteristics for their limited mobility caused by a weaker vision, hearing, strength and sensory ability. For example, the elders are 40% slower than the adults when they walk in crowded urban walkways [20] and have lower speeds when they go downstairs [21]. The fundamental diagrams and step widths of pedestrians in single-file flow show great difference between elders and young adults [22–24]. The fundamental diagram of elderly pedestrians in straight corridors was studied in [25], where it was shown that differences in the dynamics can be observed, although by means of a transformation using the desired speed, the authors could show that the fundamental diagrams of elderly pedestrians can match the fundamental diagram of younger pedestrians. Meanwhile, it was found that the elderly pedestrians have smaller interpersonal distance and bigger distance to walls. Nevertheless, the movement characteristics of elderly pedestrians through bottlenecks are still unknown. The lane formation, the relation between flow and bottleneck width as well as the speed distribution for the elders at bottleneck scenarios are necessary to be investigated. Whether the elderly movement in the bottleneck is comparable to young people also needs to be studied.

Based on these considerations, experiments under laboratory conditions of elderly pedestrians through bottlenecks were performed. The aim of this paper is to study the movement and characteristic of the elderly flow. The remaining of this paper is constructed as follows: section 2 describes the setup of the experiments. In section 3, the main analysis results including trajectories, spatial-temporal characteristics, density and speed distribution, the flow and the waiting time of the elderly crowd in bottlenecks are presented. Finally, we summarize the paper and make conclusions in section 4.

## 2. Experimental setup

The experiments were carried out in March 2018 in Hefei, Anhui province, China and 73 elderly volunteers were recruited from a senior center including 21 males and 52 females. The age of participants is 69.7±7 years old ranging from 52 to 81 and the mean height is 163 cm ranging from 150 cm to 175 cm. Fig.1 shows the distribution of age and gender of the participants. Two digital cameras (Type: HDR-SR11, resolution: 1920 x 1080 pixels, frame rate: 25 fps) were mounted on the roof of a building about 10m high to record the process of these experiments. Each participant was asked to wear a red or blue hat to ease post processing of the video recordings. The software *PeTrack* [26] was used to extract the trajectories automatically. The average height of 163 cm was used for data transformation from pixel coordinates to physical coordinates.

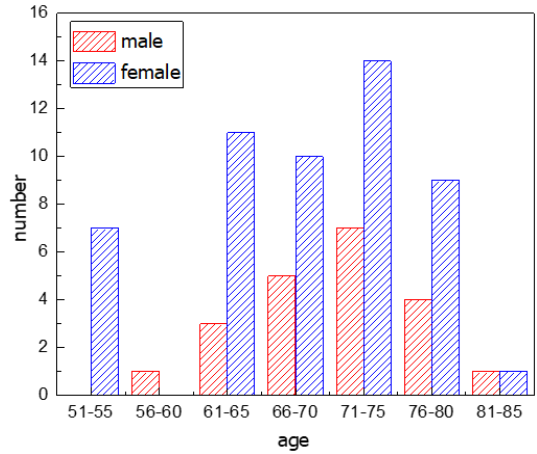

**Fig.1.** The distribution of age and gender of the pedestrians in the experiments.

Fig.2 illustrates the experimental scenario. The experimental scenario was surrounded by 1.8m high boards to construct an area of 5×3.5 m$^2$. During the analysis, we focus on the 2×2 m$^2$ measurement area in front of the exit ($x$=3~5 m, $y$=0.75~2.75 m). Initially, the participants stood in a line of seven and waited in the "Holding area". When the experiment started, they walked through the bottleneck quickly under the premise of safety.

The width of the exit $b$ was changed from 0.5 m to 1.8 m to investigate their influence on the flow. A total of 11 runs were carried out. Due to the limited physical strength of the participants, we performed each run only once. The detailed information on each run can be found in Table.1. Even though the participants had no physical problem for normal movement, we did not ask them to participate in each run by taking account of their demands for rest. This is the reason why the number of pedestrians in each run was not always identical.

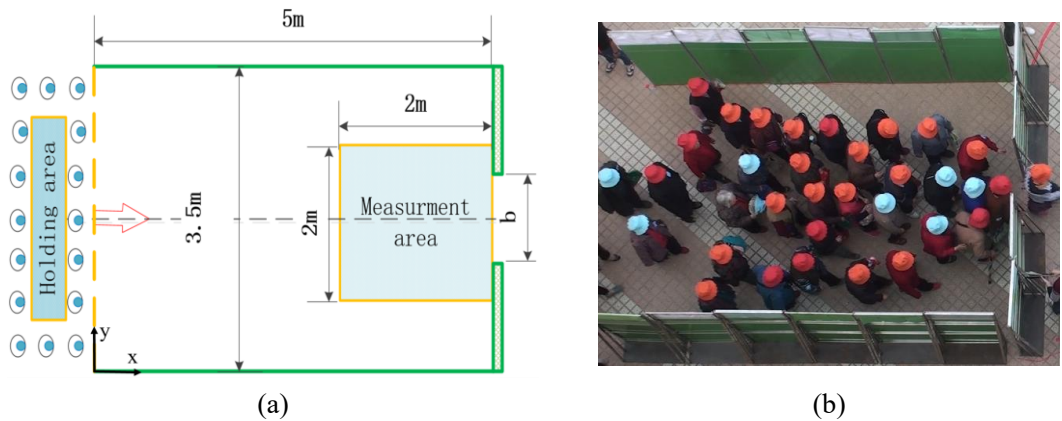

(a)  (b)

**Fig.2.** The sketch (a) and a screenshot (b) of the bottleneck experiment.

**Table. 1.** Controlled conditions and number of participants in every run.

| Index | Conditions | Number of pedestrians |
|---|---|---|
| B-050 | $b$=0.5m | 65 |
| B-060 | b=0.6m | 65 |
| B-070 | b=0.7m | 63 |
| B-080 | b=0.8m | 44 |
| B-090 | b=0.9m | 73 |
| B-100 | b=1.0m | 63 |
| B-110 | b=1.1m | 55 |
| B-120 | b=1.2m | 45 |
| B-140 | b=1.4m | 34 |
| B-160 | b=1.6m | 35 |
| B-180 | b=1.8m | 55 |

## 3. Results and Analysis

### 3.1 Trajectories

Fig.3 shows the trajectories obtained from the video recordings with instantaneous speed in different scenarios. Based on these trajectories, movement characteristics of the elders such as density, speed and flow can be calculated. The speed decreases in the process from the beginning of the movement to the exits especially in the area near the bottleneck and increases after passing the bottleneck.

In many studies it has been shown that typical arch-like formation at the exit under high density [27] appear. However in our experiment with elderly participants, the shape of the outline of the trajectories show a narrow arrow-like structure. The shape of the trajectories of young students in the recent research [28] with small exit's width is, compared to our experiment, more wide in front of the exit. This can be explained by the fact that the elderly pedestrians are extremely purposeful. They choose to wait at the back of the line to take the shortest route to the exit, with few exceptions, as in the run of B-060 (6 pedestrians) and in the run B-070 (one pedestrian), where noticeable detours were made.

Lane formation is considered as a self-organization phenomenon, often observed when pedestrians move through bottlenecks [4,12]. While in our experiment, two lanes can be observed from the trajectories of $b=0.9$ m. Deeper analysis of the spatial characteristics of pedestrian distribution at the exits will be made in section 3.2.

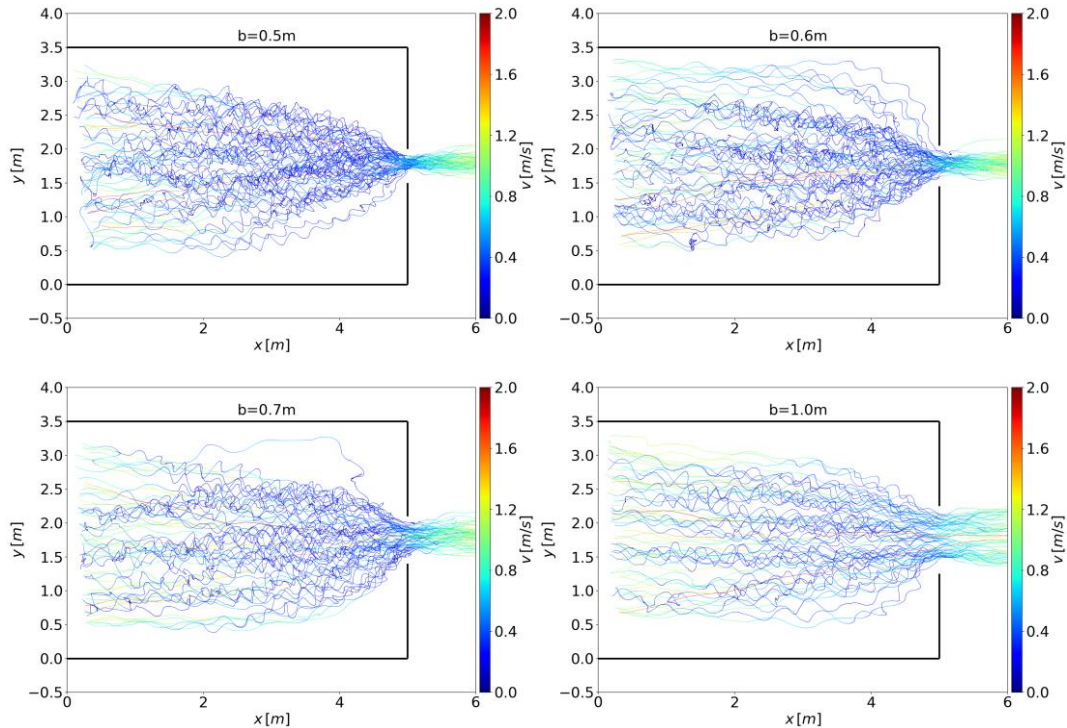

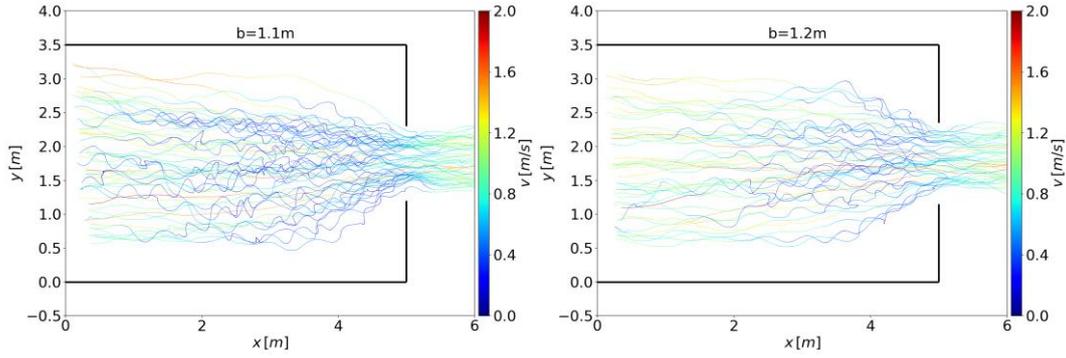

**Fig.3.** Trajectories of all elders in the experiments and the instantaneous speeds of each one are indicated by the color code. The trajectories show an arrow-like formation in front of the bottleneck and the speed decreases in front of the bottleneck, only to increase after passing the bottleneck.

### 3.2 Spatial-temporal characteristics

To investigate the influence of different exit widths on the pedestrian flow, we plot the spatiotemporal diagrams of each run by sampling the line in the exit position ($x$=5 m) of pixels from every frame and stacking them together [26]. See Fig.4. In the spatiotemporal diagrams, the axis of the direction of pedestrian movement represents the time information while the orthogonal axis represents the spatial information.

From the spatial axis, the relative position of each pedestrian at the exit can be observed. When $b$=0.5 and 0.6 m, we observe that the width can only accommodate one pedestrian, which leads to an evacuation through and they always in the middle of the exits to go through. While the passing pedestrians appear on both sides of the exit alternatively which proofs indicates the existence of the 'zipper effect' during the steady stage in the runs of $b$=0.7 and 0.8 m, but these widths still allow only one pedestrian is able to go through the exit at the same time. As a result, the probability of finding a pedestrian at position y shows a double peak structures in Fig.5 (b). When $b$=0.9 and 1.0 m, two pedestrians can pass the exit side by side. It can be seen that one and two pedestrians appear alternatively for $b$=0.9 m while, for $b$=1.0 m in most cases two pedestrians can pass side by side is the majority for $b$=1.0 m. When $b$⩾1.1 m, the exit can accommodate three or four pedestrians at the same time respectively. The probability of finding a pedestrian at position $y$ in the bottleneck exits ($x$=5.0 m) is counted for further study of the lane formation (Fig.5 (b)). The double-peak structure in the probability distribution of the positions for $b$= 0.8, 0.9, and 1.1 m display the formation of two lanes. While there are three lanes for $b$=1.2 and 1.8 m which is in accordance with the results in [12], where three lanes were formed for $b$=120 cm, $l$=6 cm.

For more quantitative analysis of the spatial characteristics at the bottleneck, the distance headway between pedestrian $i$ who is passing the exit and its closest rear neighbor is counted (Fig.5 (a)). When $b$ is smaller than 0.8 m, the distance headway decreases with the increase of the bottleneck width. This is because, as seen in the above mentioned analysis, pedestrians in narrow exits can only pass through the center of the exit, which leads to pedestrians waiting directly behind their front neighbor. In order to pass through quickly, pedestrians exhibit body overlap and cross even the 'zipper effect' as the width of the exit increases which reduces the distance headway. When $b$ is larger than 1.0 m, the distance headway increases firstly and then stabilizes as the density before the exit decreases.

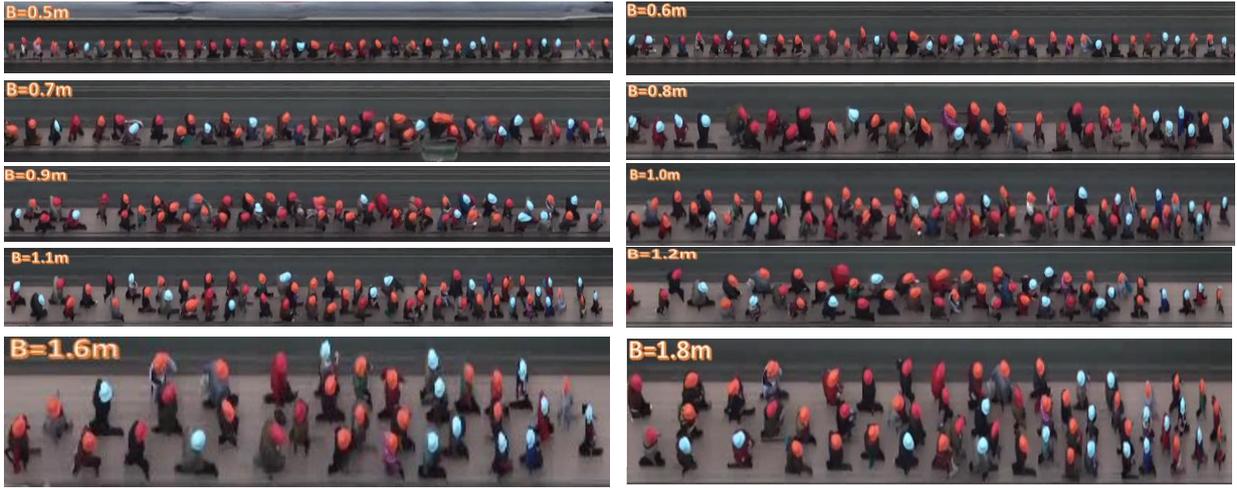

**Fig.4.** The spatiotemporal diagrams of different runs with different widths. The axis of the direction of pedestrian movement represents the time information while the orthogonal one represents the position.

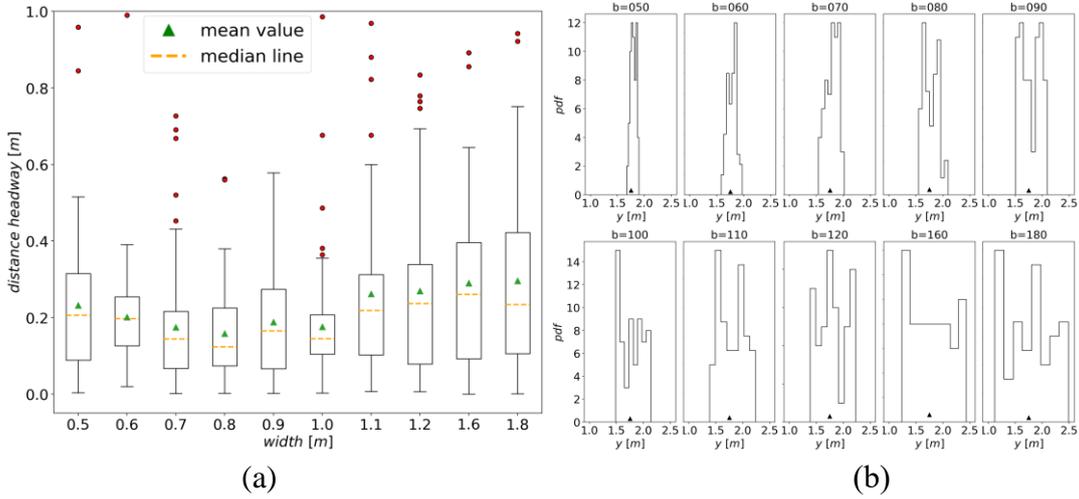

**Fig.5.** (a): The box plot distribution of the distance headway between pedestrian *i* who is passing through the exit and its rear neighbor. (b): The probability of finding a pedestrian at the exit (x=5 m) and the solid triangles represent the centers of the exits.

From Fig.4, we observe the time lapses between passages of two successive elderly pedestrians, which are represented by the length of their relative positions on the time axis. It can be observed that with increasing exit width, the time lapses among pedestrians decrease and the degree of crowd congestion decreases. Fig.6 (a) displays the quantitative statistical results that the mean value of time lapse decrease with the increase of bottleneck width. While double-peak structures, which could be caused by the zipper effect, cannot be found in the histogram distribution of the time lapses (Fig.6 (b)).

To further explore the differences in time lapses between different age groups, the time lapses are regressed over time used the method in [16]. See Fig.7 (b). An exponential relaxation of the time gaps $T_G$ is assumed: $T_G(t)=a \cdot \exp(-b \cdot t)+c$. A comparison is made between the regression results of the elderly and the young in [16] (Fig.7 (a)). It can be observed that the elderly have longer time lapses than the young under the same wide exit. The difference between the time lapses of the elderly and the young decrease with the increasing width for it is (in seconds) 0.212±0.015, 0.193±0.030, 0.132±0.020, 0.037±0.034 for *b*=0.5, 0.7, 0.8, 1.0

m respectively.

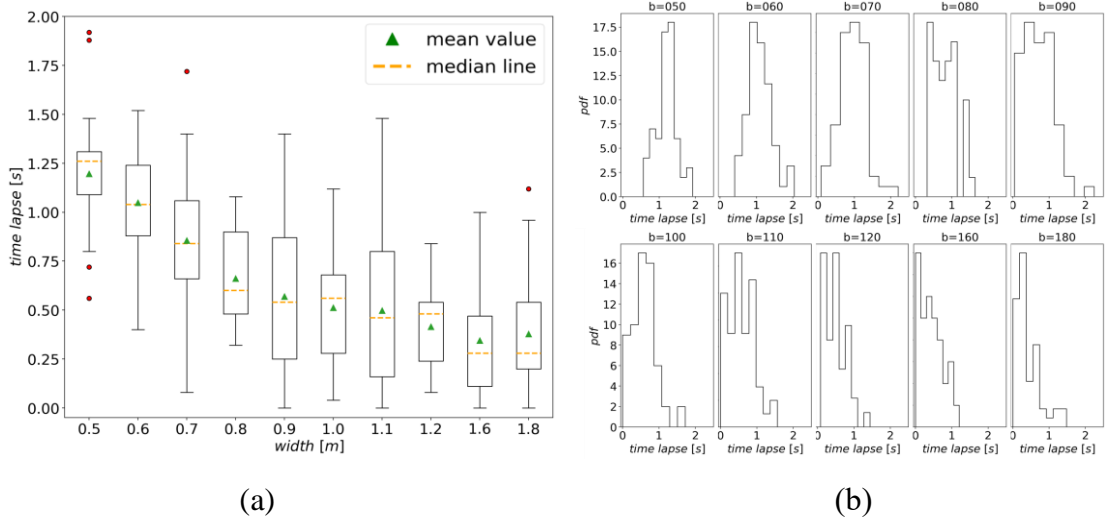

**Fig.6.** The box plot distribution (a) and the histogram distribution (b) of the time lapses between successive pedestrians passing through the exit.

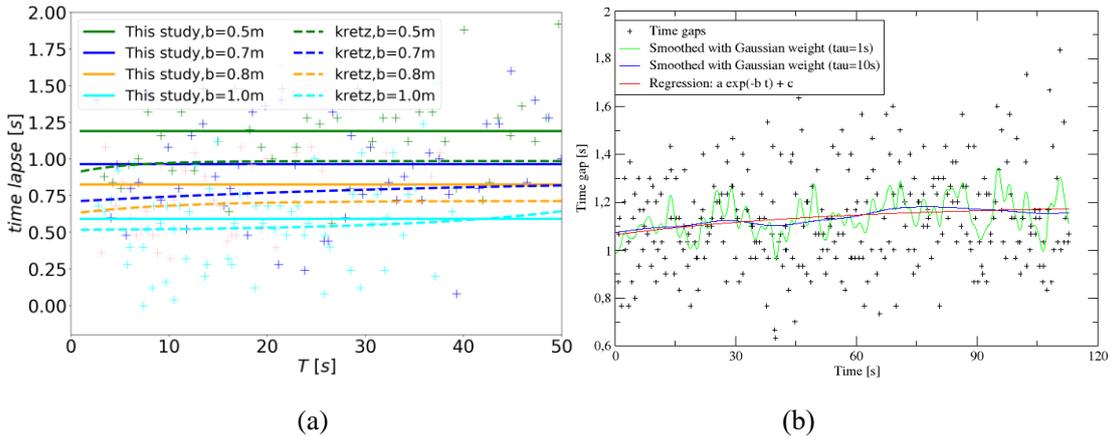

**Fig.7.** Time gaps variation with the time. (a) The original time gap ('+'s), the regression following $T_G(t)= a.\exp(-b.t)+c$ for the elderly (solid line) and the results in [16] (Kretz) with young pedestrians (dashed line). (b) The regression results in [16] for the run of exit width is 0.4 m.

For the spatial-temporal characteristics of the elderly flow through the bottlenecks, typical self-organized phenomena e.g. 'zipper effect' and 'lane formation' are observed from the spatiotemporal diagrams. The elderly pedestrians have 0.212±0.015 s longer time lapse than the young when passing through the bottleneck.

### 3.3 Density and speed distribution

To quantitatively investigate the dynamics of elderly and quantify its characteristics under different bottleneck widths, we calculate the time series of density and speed in the measurement area (Fig.2 (a)) based on Voronoi method [26]. See Fig.8.

The mean Voronoi density firstly increases, then stabilizes and finally decreases. The density shows a relatively short plateau, which is due to the limited number of participants. Therefore, the steady state [29] at the measurement area is selected mainly according to the mean speed, which showcases a relatively more stable trend. The start and the end of the steady state are represented by the vertical dashed lines in Fig.8.

When the width is 0.7 m and 0.9 m, the mean density can reach 3.4 m$^{-2}$, which is lower than 5 m$^{-2}$ of young adults in the previous experiments [6]. We hypothesized that the lower mobility of elders leads to reduced density before the bottleneck. The mean Voronoi speed decreases with increasing number of pedestrians in the measurement area, it stabilizes when the number of pedestrians in the measurement area remain unchanged. When $b \leq 1.1$ m the speed stabilizes around 0.5 m/s for the limitation of the congestion. When $b \geq 1.2$ m the speed increases as the increasing width due to the waning impact of congestion. In the study of pedestrian safety evacuation, we focus on the characteristics of pedestrian movement in the congestion process.

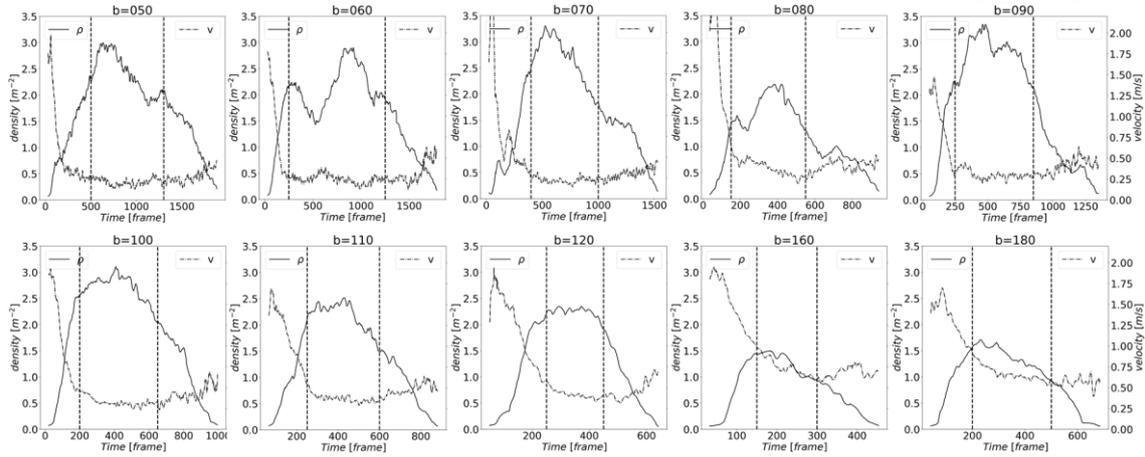

**Fig.8.** Time series of density and speed in the measurement area of each run except for B-350-350-140 which has the fewest participants. The two vertical lines represent the start and end of the relatively steady state.

For further study of the density and speed characteristics of the elderly crowd in the bottleneck scenario, Fig.9 and 10 show the spatial distribution of the mean density and speed in steady state respectively. For scenarios with different exits widths, the distribution characteristics of density are similar. The pick density is at the position of $x=3\sim4$ m and the density decreases at the position of exit $x=5$ m. The highest local density is beyond 4 m$^{-2}$ in run of $b=0.5$ m. For $b=1.8$ m, the highest density is near the exit and the density is more uniform in the whole scenario compared with the narrower widths. Since in $b=1.8$ m three or four pedestrians can go through the exit at the same time leading to the increasing density in the exit.

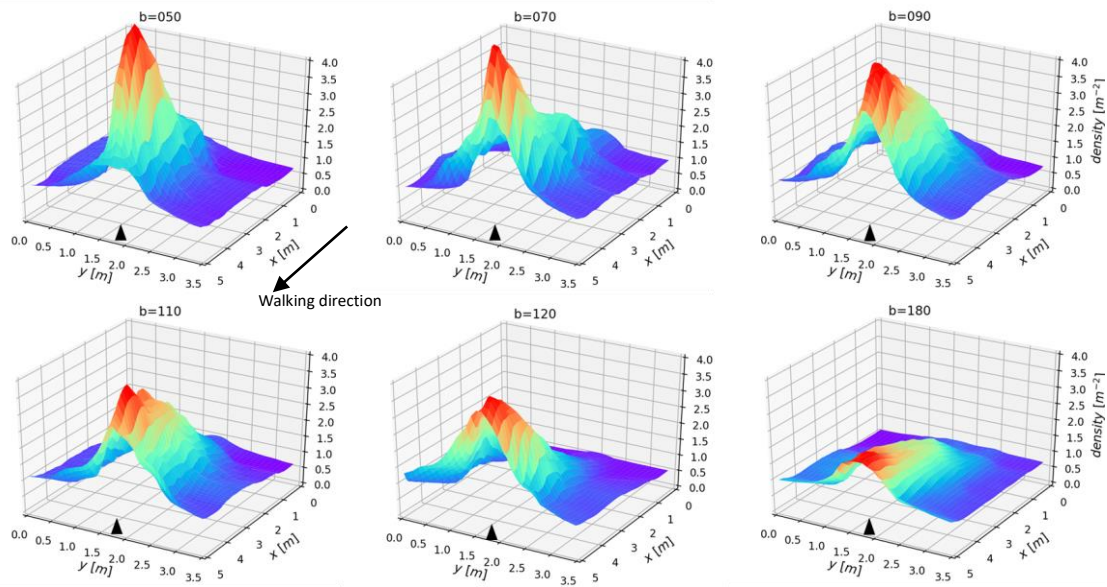

**Fig.9.** The spatial distribution of density under different widths. The line with an arrow indicates the walking direction. The solid triangle represents the center of the exit.

For the distribution of mean speed in the steady state (Fig.10), the speed increases with increasing width. While in $b$=1.2 m fewer pedestrians than $b$=1.8 m exist, which leads to higher speeds in $b$=1.2 m than in b=1.8 m.. The speed of pedestrians in both sides of the walking direction is higher than in the middle. In runs $b$=0.9 and 1.1 m, the speed on the left and right sides is not symmetrical, due to some detours. It can be observed that the elderly pedestrians have higher speed at the beginning of the movement ($x$=0~2 m). Then, the speed decreases in the middle area of the scenario ($x$=2~5 m). Finally, the pedestrians passing through the exit with higher speed. For $b$=1.8 m, there is no obvious change of the speed in the scene. It implies that the restriction of exit width on pedestrian flow is reduced. The speed in the middle is higher than that in both sides of the walking direction. This can be explained by the tendency of pedestrians to walk through the middle paths of the bottleneck.

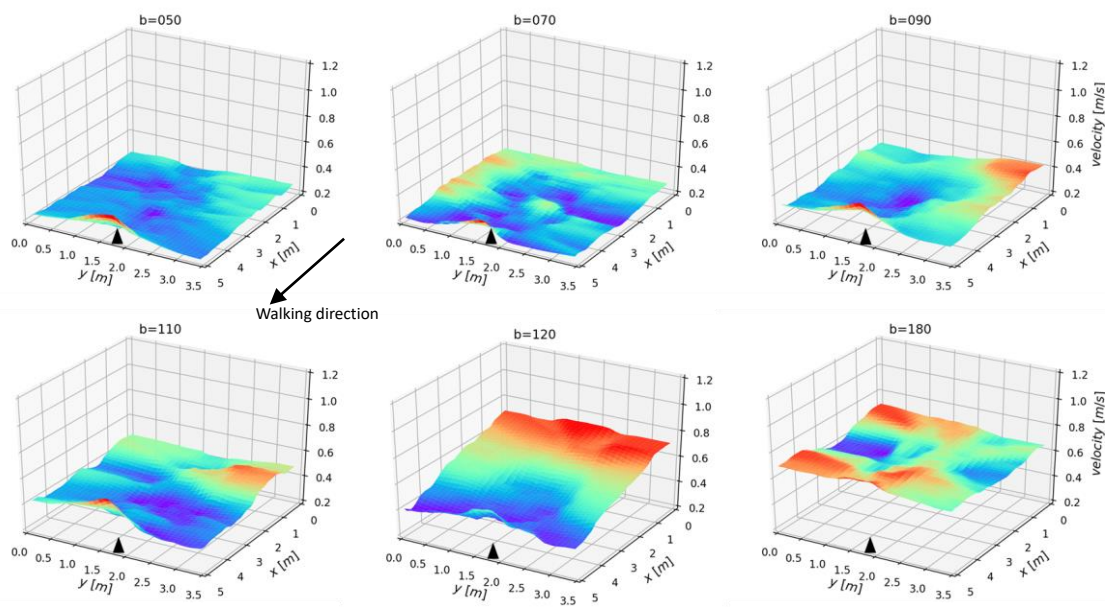

**Fig.10.** The spatial distribution of speed under different widths. The line with an arrow indicates the walking

direction. The solid triangle represents the center of the exit.

In this section, we studied the spatial distribution of the density and speed of the elderly flow during the steady stage. When $b<1.1$ m, the density reaches the peak at the position of $x=3\sim4$ m while it decreases at the bottleneck around $x=5$ m. When $b<1.2$ m, the speed in the middle is lower than that in both sides of the walking direction but the opposite is true for $b=1.8$m.

**3.4 Flow of the elderly pedestrians in bottlenecks**

Considering the inconsistency of the number of pedestrians in each run, we select the first 30 pedestrians in front of the line in each run and calculate the evacuation time accordingly. From Fig.11 (a) we can see that the total evacuation time decrease with the bottleneck width. The slope of the relationship between evacuation time and width of exits decreases around $b=1.1$ m, since at this width three pedestrians to go through the exit at the same time. However, we cannot exclude the factor of the number of pedestrians.

The relationship between the flow $J$ and the width $b$ is studied in Fig.11 (b). The flow depends on the number of pedestrians and the density around the exit. Considering that the number of pedestrians in each run is different, only elders passing through the bottleneck during the steady-state stage (Fig.8) are considered. The flow increases linearly with the width $b$. We only fit the data of $b<1.2$m considering that there are not enough pedestrians in the runs of $b=1.4$ and 1.6 m:

$$J=2.18b-0.30 \text{ [s}^{-1}\text{]}.$$

In [6] it was shown that the restriction effect of a short narrowing on pedestrian flow is smaller than long narrowing. Therefore, for better comparability we use the data from [4,16] whose experimental scenarios are bottlenecks with short narrowing. From Fig.11, it can be observed that the elderly have lower flow than the young in [4] (Rupprecht), while the difference between the elderly in our study and the young in [16] (Kretz) is small, especially at low exit's widths.

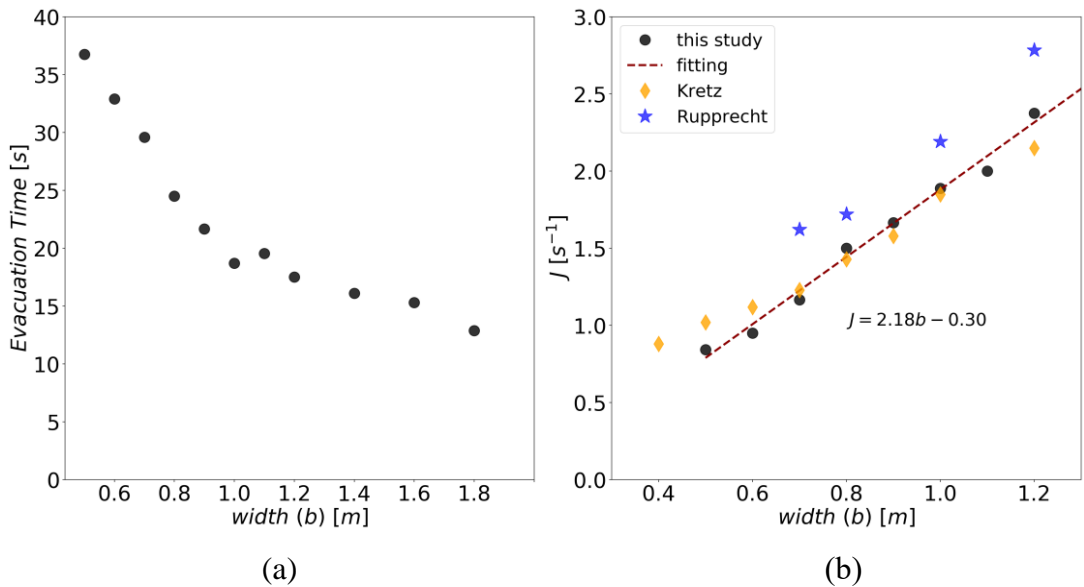

(a)　　　　　　　　　　　　　(b)

**Fig.11.** The relation between the total evacuation time of the first 30 pedestrians and bottleneck width (a), the cumulative evacuees and the cumulative time of the evacuation (b).

For further study of the blockages in front of bottlenecks, we investigate the changes in the time lapses τ based on the Clauset-Shalizi-Newman method [30]. The data of the first 5 pedestrians and the last 10 pedestrians in each run were removed, since they represent a transient state. Furthermore, the CCDF is fitted as

$$p(\tau) \sim \tau^{-\alpha}$$

with $x_{\min}$=0.5. We found that the exponent $\alpha$ shows a strong positive correlation to the bottleneck's width. See Table 2. It implies that the narrower exits have greater possibility of having longer clogs.

**Table. 2.** The exponent $\alpha$ obtained from the power-law analysis with $x_{\min}$=0.5.

| Width $b$ (m) | $\alpha$ | Width $b$ (m) | $\alpha$ |
|---|---|---|---|
| 0.5 | 2.66 | 0.9 | 3.71 |
| 0.6 | 3.14 | 1.0 | 4.02 |
| 0.7 | 3.16 | 1.1 | 4.03 |
| 0.8 | 3.68 | 1.2 | 4.33 |

In this section, we studied that the flow of the elderly through the bottlenecks. The flow of the elderly is lower than that of the young under the same exit width. The blockages and the evacuation time increase with the bottleneck's width. The exponent α shows a strong positive correlation to the bottleneck's width.

### 3.5 Waiting time

Waiting time in bottlenecks may be an indicator to distinguish between queuing and pushing behavior. From the analysis in [18,19] with young participants, we know that two phases can be observed in the diagrams. The motivation and the width of the corridor have obvious influence on the slopes and duration of different phases. Note that we cannot draw any conclusion from a direct comparison with [18,19], since the geometrical setup of the experiments is different. Based on the above considerations, we compare the waiting time of the elderly under different exit width. In this analysis, only trajectories data after the beginning of the steady stage are used. For each pedestrian, the blue dots represent the initial position (Fig.12). The initial distance is the Euclidean distance between the position the pedestrian when she/he first appears during the measurement time to the target, which is the position chosen to passing through the entrance. The initial time to the entrance is given by the total time of the pedestrian needed to arrive to the target. Then, the distance and time to the target of each one is calculated every 10 frames of its movement time during the measurement time.

Two phases with different slopes can be observed for $b$=0.5, 0.6 and 0.7 m (Fig.12 shows the result of width $b$=0.6 m). We observe a strong reduction of distance to the target and a small reduction of the time to the target in the contraction phase. For the waiting phase in the red circle, the tracked positions show a stronger reduction of the time to the target with smaller reduction of the distance to the target. The pedestrians in the waiting phase can only move forward when their neighbors in front are moving. The coexistence of two phases gradually disappears as the width increases. Only the contraction phase can be observed when the width $b$>0.8 m. This indicates that narrow exit widths block the waiting phase and lead to a longer contraction phase, since the slopes of these tracked position lines increase. The slopes of the tracked lines decrease with the increasing width which means the increased capacity and reduced congestion levels.

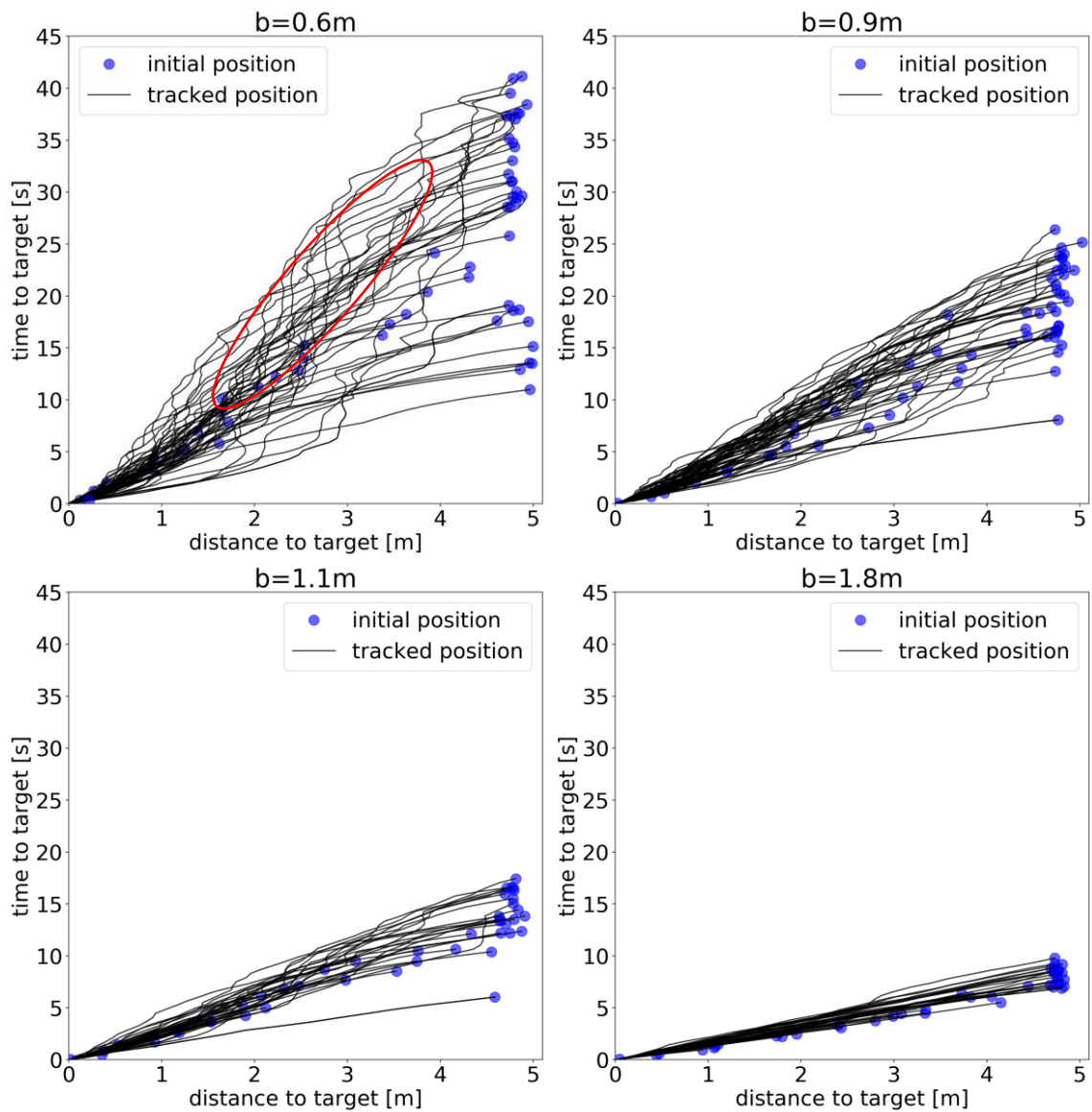

**Fig.12.** Waiting time versus linear distance to the target in the exit. The blue dots represent the initial distance and total time from the first appeared position to the target during the measurement time. A time interval of 10 frames is used to calculate the tracked position for each elderly pedestrian.

## 4. Summary

In this work, we experimentally study the evacuation characteristics of elderly people through a bottleneck. 73 elderly volunteers with the age ranging from 52 to 81 were recruited from a senior center. 11 runs experiment with different bottleneck width from 0.5 m to 1.8 m were carried out. Most of the elderly pedestrians chose the shortest path from the exit.

The spatial-temporal characteristics can be observed from the spatiotemporal diagrams and the distance headways and time lapses were counted to make further quantitative analysis. We could observe the zipper-effect, meaning pedestrians appear on both sides of the exit alternatively in the steady state of runs with $b$=0.7 m and 0.8 m. The width of $b$=0.9 m, $b$=1.1 m and $b$=1.8 m can accommodate two, three and four elderly pedestrians at the same time respectively. The distance headway first decreases and then increases with the increasing width. For the probability of finding a pedestrian at position y in the bottleneck exits ($x$=5.0 m), there are two double peaks in runs with $b$=0.8, 0.9 and 1.1 m and three peaks in runs with $b$=1.2 and 1.8 m. The time lapse between two consecutive pedestrians decreases with the width. The elderly have around 0.2 s longer time lapse when passing through the bottlenecks of $b$=0.5 and 0.7 m compared with the young pedestrians. And the difference is dependent on the width for it decreases as the width increase.

The mean density and speed vary over the whole space and shows great dependency on the location in the scenario. The peak density appears at x=3~4 m in front of the exits. The highest local density is beyond 4 $m^{-2}$. After analysing the speed spatial distribution, we observe that pedestrians first slow down and then accelerate to through the scene. Finally, they pass through the exits with a higher speed. The speeds in the both sides of the walking direction are higher than the middle and the left side with detour is higher than the right.

The total evacuation time of the first 30 pedestrians displays a piecewise linear relation with the exit width, with $b$=1.0 m being the intersection point.

It takes less time to evacuate the same number of people and more people are evacuated within the same time with the increment of width. Furthermore, the flow $J$ and the CCDF of time lapse $τ$ were analyzed to evaluate the clogs of different wide bottlenecks. Compared with the young in [16], the elderly have lower flow $J$ for the same exit width. The specific flow is higher in front of the exit than that in the beginning area. The area with high specific flow increases and moves backward with the increasing width. The waiting times of elderly pedestrians under different wide bottleneck were analyzed. Two phases including waiting phase and contraction phase can be observed when the width $b$=0.5, 0.6 and 0.7m.

The analysis in this paper enriches the empirical data of the elderly crowd, especially in bottleneck scenarios. It is helpful for the modeling and the design of facilities for the elderly population. Further studies should be carried out to study the inherent difference between elders and young adults under the same conditions by means of experiments as well as numerical simulations.

## Acknowledgements

The authors acknowledge the foundation support from the National Natural Science Foundation of China (Grant No. 71704168), from Anhui Provincial Natural Science Foundation (Grant No.1808085MG217) and the Fundamental Research Funds for the Central Universities (Grant No. WK2320000040).